\documentstyle[preprint,aps]{revtex}

\begin{document}
\draft

\title{
Charged Leptons with Nanosecond Lifetimes}

\author{Marc Sher}

\address{Physics Department, College of William and Mary,
Williamsburg, VA  23187,  USA}

\date{\today}

\maketitle

  \begin{abstract}
 Some extensions of the standard model contain additional leptons which
are vectorlike under weak isospin.  A class of models is considered in which
these
leptons do not appreciably mix with the known leptons.  In such models, the
heavy
charged lepton and the heavy neutrino are degenerate in mass, and the
degeneracy is
broken by radiative corrections.  The mass splitting is calculated and found to
be
very weakly dependent on the lepton mass, varying from 250 to 330 MeV as the
mass
varies from 100 to 800 GeV.  This result is {\it not} affected significantly by
inclusion in a supersymmetric model in spite of the additional loops involving
the
superpartners.  As a result, this fairly general class of models has a charged
lepton
whose lifetime varies in the narrow range from 0.5 to 2.0 nanoseconds, and
which decays
into neutrals plus a very low energy electron or muon.
\end{abstract}
\centerline{submitted to Brief Reports}

\newpage

The relatively recent discovery that only three light neutrinos exist has
 constrained  possible extensions of the standard model.  It is no longer
possible to
add a  fourth sequential lepton family (with a massless neutrino) to the model.
Extensions which do add another generation of fermions tend to fall into one of
three
types:  (a) sequential fermions with a right-handed neutrino, (b) mirror
fermions,
popular in left-right symmetric models,  and (c) vectorlike fermions.
  Vectorlike
fermions are an essential ingredient in the aspon model\cite{aspon}, which
gauges the
Peccei-Quinn symmetry to solve the strong CP problem, and vectorlike leptons
appear in
superstring-inspired $E_6$ models\cite{others}.

Although the phenomenology of all of the various types of heavy fermions has
been
extensively studied\cite{dpf}, in this Brief Report I will note a feature
present in
some models with vectorlike fermions that has not (to my knowledge) been
discussed, and
which will have some precise and unusual phenomenological implications. The
focus will
be on the leptonic sector, although comments will briefly be made on  the
extension to
the quark sector. The specific models under consideration will be  those in
which the
vectorlike leptons do not appreciably mix with the lighter leptons.  The
suppression
of such mixing can easily be arranged via conservation of $e$, $\mu$ and
$\tau$-number, which can be imposed via a discrete symmetry. Alternatively, one
can
suppose that any mixing that does exist comes from the  neutrino sector, and
thus the
mixing angle would be of the order of the $\nu_\tau$  mass to the vectorlike
neutrino
mass, which (using the cosmological bound on the tau  neutrino mass) must be
less than
$10^{-10}$\cite{yuan}.

In such models, the vectorlike leptons, $\left(N\atop L\right)_L$ and
$\left(N\atop L\right)_R$, have no couplings to Higgs doublets, and are
degenerate
in mass.  This degeneracy will be split by radiative corrections.  The mass
splitting
will be calculated here, and the ``unusual" feature is that the mass splitting
is
extremely insensitive to the lepton mass, and thus the lifetime of the $L$ will
be in
the  very narrow range of $0.5-2.0$ nanoseconds, which is certainly
phenomenologically
 interesting.  Even if one extends the model to include supersymmetry (which
generally
 can significantly change radiative corrections), this mass splitting and
lifetime are
 {\it not} significantly changed.  The splitting and lifetime will be
calculated, and
the  phenomenological implications discussed.

In the simplest case, the $L$ and $N$ are degenerate in mass at tree level,
with mass
$M$, and this degeneracy is broken by the radiative corrections shown in Fig.
1.
Contributions from charged vector bosons will cancel in the mass difference.
The mass
splitting is easily calculated to be
$$\Delta M\equiv M_L-M_N={\alpha M \over 2\pi}\int_0^1\ dx\ (1+x)\left(\ln(x^2+
{m^2_Z\over M^2}(1-x))-\ln(x^2)\right)$$
Note that this expression will vanish in the limit as $m_Z\rightarrow 0$.  This
is
expected, since in that limit the SU(2) gauge symmetry is unbroken and the
$L$-$N$
mass degeneracy is thus unbroken.  This fact is crucial, and is responsible for
the
relatively mild $M$-dependence of the mass-splitting.  The charged lepton $L$
is
always heavier (this is reassuring since new heavy charged stable particles are
cosmologically excluded); and the above expression is plotted as a function of
$M$ in
Fig. 2.

The $L$ will decay via $L\rightarrow Ne\overline{\nu}$, $L\rightarrow
N\mu\overline
{\nu}$ (decays into pions, possible for the heavier masses, will be phase-space
suppressed).  The decay width into $Ne\overline{\nu}$ is given by (note that
the
coupling to the W is vectorlike)
$$\Gamma={g^2(\Delta M)^5\over 960\pi^3 m_W^4}$$ where $g$ is the electroweak
coupling.  The decay into $N\mu\overline{\nu}$ will have an identical width
with a
mild phase-space suppression.  The total lifetime is plotted as well in Fig. 2.

The lifetime of  $0.5-2.0$ nanoseconds is very interesting phenomenologically.
 One would see a charged particle travel about a meter, then decay into a low
energy
(a few hundred MeV) electron or muon plus missing energy.  Such a signal should
have
a relatively low background and may allow detection of such a lepton up to much
higher
 masses than conventional sequential leptons (for which backgrounds, in a
hadron
collider, are serious problems).  For vectorlike quarks, the results are
identical,
with an additional factor of $1/3$ due to the quark charges.
This leads to lifetimes of $2.5-10$ microseconds, thus only a fraction of a
percent
or so will decay (through $U\rightarrow De\overline{\nu}$) in the detector.
 Still,
the higher production rate at a hadron collider makes the detection quite
likely (for
 example, the Tevatron has published limits on ``stable" quarks, but not
 ``stable"
leptons\cite{tev}).

How model-dependent is this result?  The only assumption has been the existence
of
vectorlike leptons which do not appreciably mix with the lighter leptons.  If
one
adds additional Higgs doublets and singlets, the result will not change.
 Adding
additional vectorlike leptons will also not change the result, nor will adding
additional gauge groups (as long as the left and right-handed doublets
transform
identically under these groups).  The most popular extension of the standard
model is
supersymmetry; we now consider the mass splitting in supersymmetric models.

In supersymmetric
 models, one must also consider the diagrams of Fig. 3, in which the
$\tilde{L}$ and
$\tilde{N}$ are the superpartners of the $L$ and $N$, and where $\chi_j$ are
the
neutralinos (as with the $W$'s, charginos will not contribute in the mass
difference).
  Although the $L$ and $N$ do not couple to Higgs fields, they could couple to
Higgsinos due to gaugino-Higgsino mixing, and thus all four neutralinos must be
considered.  The Feynman rules can be extracted from the work of Haber and
Kane\cite{susy}.   For example, the vertex in which an $L$ goes into a
$\tilde{L}_L$
and a $\chi_j$ is  given by $-i{g\over\sqrt{2}}\sec\theta_W(1-\gamma_{
_5})(-{1\over
2}+\sin^2\theta_W)N_{j2}+i{e\over\sqrt{2}}(1-\gamma_{ _5})N_{j1}$, where $N$ is
the
matrix which diagonalizes the neutralino mass matrix.  Care must be taken to
use
Majorana fermion propagators for the neutralinos.

The resulting mass splitting is given by
$${\alpha\over 2\pi}M\sum_j A_j
\int_0^1dx\ (1-x)\ln[\tilde{m}_L^2(1-x)+m^2_{\chi_j}x-M^2x(1-x)]$$
where $$A_j\equiv \left(|N_{j1}|^2-|N_{j2}|^2+\cot\ 2\theta_w(N_{j2}^*
N_{j1}+N_{j1}^*N_{j2})\right)$$ and $N$ diagonalizes the mass matrix
$$\pmatrix{M_2\sin^2\theta_w+M_1\cos^2\theta_w&(M_2-M_1)\cos\theta_w\sin\theta_w&0&0
\cr (M_2-M_1)\cos\theta_w\sin\theta_w&
M_2\cos^2\theta_w+M_1\sin^2\theta_w&m_Z&0\cr 0&m_Z&\mu\sin 2\beta&
\mu\cos 2\beta\cr 0&0&\mu\cos 2\beta&-\mu\sin 2\beta\cr}$$
Here, $M_1={5\over 3}\tan^2\theta_wM_2$, $M_2$ is the SU(2) gaugino mass
parameter, $\mu$ is the supersymmetric mass parameter and $\tan\beta$ is the
ratio of
vacuum expectation values.

Note that the divergences in the diagrams can be obtained by replacing the
integral
with a ${1\over\epsilon}$, and they cancel since $|N_{j1}|^2=|N_{j2}|^2=1$ for
a
unitary matrix and the individual columns are orthogonal.  It is interesting to
see
how the above result vanishes in the limit that $m_Z\rightarrow 0$, as it must.
 In
that case, the mass matrix divides into two $2\times 2$ matrices, and only the
$j=1,2$
 terms contribute.  The upper $2\times 2$ matrix can be trivially
diagonalized---the
mixing angle is given by the negative of the Weinberg angle.   Thus,
$N_{11}=N_{22}=
\cos\theta_w$ and $N_{21}=-N_{12}=\sin\theta_w$.  Plugging into the above, for
{\it
each} term in the sum over $j$, the coefficient of the integral vanishes.

Thus, the result has the expected properties, yet it appears to depend on
several
unknown parameters.  However, it turns out that, for reasonable values of these
parameters, the numerical contribution is extremely small.  To see why,
consider
the limit of the result when $m_Z^2<<|M_2-\mu|^2$, which is the case for much
of the
parameter space.  In this case, one can use standard quantum mechanical
time-independent perturbation theory; the small expansion parameter is
$\lambda\equiv
{m_Z\cos\phi\cos\theta_w\over M_2-\mu}$ where $\phi\equiv {\pi\over 4}-\beta$.
The correction to the $m_Z\rightarrow 0$ limit to first order in $\lambda$
vanishes;
the second order correction arises from the normalization of the first order
``wavefunctions".
The coefficient of the integral in the above expression (for $j=1$, for
example)
becomes $$\cos^2\theta_w-{\sin^2\theta_w\over 1+\lambda^2}-2\cot\
2\theta_w{\cos\
\theta_w\sin \theta_w\over \sqrt{1+\lambda^2}}.$$  Expanding this out to
leading
order in $\lambda^2$ gives $2\lambda^2(\sin^2\theta_w-{1\over 4})$.  A similar
expression occurs for the other terms in the sum. The appearence of the
$\sin^2\theta_w-{1\over 4}$ factor (which appears unrelated to the vector
coupling of leptons to the Z in the standard model)
makes the final contribution at most a few MeV, and thus negligible.

Thus, the result that vectorlike leptons that do not mix appreciably with
lighter
leptons have a lifetime between $0.5$ and $2.0$ nanoseconds is very robust, and
is
not affected even by the addition of supersymmetry.  The generality of such
models is
sufficient that experimenters searching for exotic leptons are encouraged to
pay
particular attention to charged leptons with lifetimes of about a nanosecond
and
which decay into neutrals plus a low energy electron and/or muon.

I thank Sergei Ananyan and Yao Yuan for checking the calculation of the $L$
lifetime
in the non-supersymmetric case, and Jose Goity for useful discussions.  This
work was
supported by the National Science Foundation.

\newpage
\def\prd#1#2#3{{\it Phys. ~Rev. ~}{\bf D#1}, #3 (19#2) }
\def\plb#1#2#3{{\it Phys. ~Lett. ~}{\bf B#1}, #3 (19#2)}
\def\npb#1#2#3{{\it Nucl. ~Phys. ~}{\bf B#1}, #3 (19#2) }
\def\prl#1#2#3{{\it Phys. ~Rev. ~Lett. ~}{\bf #1}, #3 (19#2) }

\bibliographystyle{unsrt}

\newpage
\begin{figure}
\caption{Contributions to the mass splitting of a vectorlike
lepton doublet.  The masses of the $L$ and $N$ are degenerate
at tree level.}
\end{figure}

\begin{figure}
\caption{The mass splitting of a vectorlike doublet, in MeV,
as a function of the lepton doublet mass.  The lifetime is
also given in nanoseconds.}
\end{figure}

\begin{figure}
\caption{Contributions to the mass splitting in a supersymmetric
model.  $\tilde{L}$ and $\tilde{N}$ are the scalar partners of
the $L$ and $N$, and are also degenerate in mass at tree level;
$\chi_j$ is the neutralino (j=1,4).}
\end{figure}
\end{document}